# Nature of the enigmatic pseudogap state: novel magnetic order in superconducting HgBa$_2$CuO$_{4+\delta}$

Y. Li[1], V. Balédent[2], N. Barišić[3,4], Y. Cho[3,5], B. Fauqué[2], Y. Sidis[2], G. Yu[1], X. Zhao[3,6], P. Bourges[2], M. Greven[3,7]

The nature of the enigmatic pseudogap region of the phase diagram is the most important and intriguing unsolved puzzle in the field of high transition-temperature ($T_c$) superconductivity. This region, the temperature range above $T_c$ and below a characteristic temperature $T^*$, is characterized by highly anomalous magnetic, charge transport, thermodynamic and optical properties [1-2]. Associated with the pseudogap puzzle are open questions pertaining to the number of distinct phases and the presence of a quantum-critical point underneath the superconducting dome [3-5]. Here we use polarized neutron diffraction to demonstrate for the model superconductor HgBa$_2$CuO$_{4+\delta}$ (Hg1201) that $T^*$ marks the onset of an unusual magnetic order, and hence a novel state of matter with broken time-reversal symmetry. Together with prior results for YBa$_2$Cu$_3$O$_{6+\delta}$ (YBCO) [6,7], this observation constitutes an essential and decisive demonstration of the universal existence of such a state. The new findings appear to rule out a large class of theories that regard $T^*$ as a crossover temperature [8-10] rather than a phase transition temperature [11-13]. Instead, they are consistent with a variant of previously proposed charge-current-loop order [11,12] that involves apical oxygen orbitals [14], and with the notion that many of the unusual properties arise from the presence of a quantum-critical point.

[1] *Department of Physics, Stanford University, Stanford, California 94305, USA*

[2] *Laboratoire Léon Brillouin, CEA-CNRS, CEA-Saclay, 91191 Gif sur Yvette, France*

[3] *Stanford Synchrotron Radiation Laboratory, Stanford, California 94309, USA*

[4] *1. Physikalisches Institut, Universität Stuttgart, Pfaffenwaldring 57, 70550 Stuttgart, Germany*

[5] *Research Center for Dielectric and Advanced Matter Physics, Pusan National University, Busan 609-735, Korea*

[6] *State Key Lab of Inorganic Syntesis & Preparative Chemistry, College of Chemistry, Jilin University, 2699 Qianjin Street, Changchun 130012, P.R. China*

[7] *Department of Applied Physics, Stanford University, Stanford, California 94305, USA*

The correct physical picture for the high-$T_c$ phase diagram (Fig. 1a) has remained unclear, in part because there has not been conclusive experimental evidence revealing a new type of order. Neutron scattering measurements, which were primarily carried out on YBCO and only surveyed a limited part of reciprocal space, resulted in contradictory claims of no [15,16] and weak [17] magnetic order, while the interpretation of muon spin relaxation measurements on YBCO [18,19] and of circularly polarized photoemission experiments on the material $Bi_2Sr_2CaCu_2O_{8+\delta}$ [20,21] has been controversial. However, recent polarized-neutron diffraction experiments on YBCO [6,7] point to the possible existence of novel magnetic order below $T^*$ that does not break translational invariance. The observed effect occurs at positions in reciprocal space that had not been considered in prior experiments.

YBCO possesses a relatively complicated orthorhombically distorted crystal structure, with two $CuO_2$ layers forming a double-layer in the unit cell and with additional Cu-O chains between the double-layers. In order to assess if this effect is unique to YBCO or a universal property of the high-$T_c$ superconductors, it is therefore of utmost importance to extend the investigation to additional, structurally simpler compounds. Hg1201 possesses a simple tetragonal crystal structure (Fig. 1e), with only one $CuO_2$ layer in the unit cell, and the highest maximum $T_c$ of all known single-layer compounds [22,23]. These properties, together with a wide accessible doping range and minimal effects of disorder [23,24], render Hg1201 an ideal system for the clarification of the pseudogap physics. Through a recent breakthrough in crystal growth, sizable high-quality single crystals have finally become available [25], enabling the present study.

Polarized neutron diffraction experiments were performed on the 4F1 triple-axis spectrometer at the Laboratoire Léon Brillouin, Saclay, France. The experimental setup was similar to that described in [6], enabling the detection of scattered neutrons in both spin-flip (SF) and non-spin-flip (NSF) channels. We define the flipping ratio $FR=I_{NSF}/I_{SF}$ ($I$=intensity) to quantify the experimental efficiency of polarization. With careful arrangement, a stable $FR$ as high as 95 can be obtained, which proved to be crucial for the detection of small magnetic signals in samples with relatively high carrier-concentration. All measurements were performed in the ($H0L$) scattering plane, where the scattering wave vector is quoted as $\mathbf{Q} = H\,\mathbf{a^*} + K\,\mathbf{b^*} + L\,\mathbf{c^*} \equiv (HKL)$ in units of the reciprocal lattice vectors, with typical room temperature values $a^* = b^* = 1.614$ Å$^{-1}$ and $c^* = 0.657$ Å$^{-1}$. Four underdoped samples were investigated; as revealed in Fig. 1b, they exhibit sharp superconducting transitions with $T_c = 61, 79, 81,$ and 89 K.

Figure 2a-c demonstrates the existence of a magnetic component in the SF channel for samples A and B. Due to the relatively strong intensity from unavoidable nuclear Bragg peak leakage in the SF channel, the measurement was performed at the weak nuclear reflection $\mathbf{Q}$=(101). The neutron polarization was parallel to the momentum transfer, $\mathbf{P}//\mathbf{Q}$, a geometry in which all magnetic scattering occurs in the SF channel. The linear slope of the nuclear scattering observed in the NSF channel can be accounted for by the Debye-Waller factor. As expected, due to the leakage, the SF data exhibit a linear nuclear scattering contribution as well. However, the SF data furthermore exhibit an additional component below $T_{mag}$~250 K, which we conclude to be of magnetic origin (see also below). The two samples have nearly identical values of $T_c$ and $T_{mag}$, and the strength of the magnetic signal is nearly indistinguishable after normalization by the nuclear scattering intensity (Fig. 2b,c). The onset of magnetic order in YBCO has been associated with the pseudogap temperature $T^*_\rho$ determined by resistivity measurements [6]. Resistivity data for a separate small crystal with $T_c = 79$ K are shown in Fig. 1c. Indeed, the rescaled magnetic intensities for samples A and B follow quite well the deviation from linear resistivity (Fig. 1d), strongly suggesting that the observed magnetic and charge properties share the same physical origin.

For novel magnetic order associated with the pseudogap phase, it is expected that the ordering temperature and strength increase (decrease) toward lower (higher) doping. In order to test this, we subjected sample B to a reducing heat treatment that lowered the oxygen (and, consequently, the hole carrier) concentration. The resultant sample B' exhibits a significantly lower $T_c$ of 61 K. Indeed, as displayed in Fig. 2e,f, the onset temperature of the magnetic order has shifted to significantly higher temperature, and the strength of signal has increased by more than a factor of five. On the other hand, for the most highly doped crystal (sample C; $T_c = 89$ K), we were no longer able to discern a magnetic signal within the counting statistics of the experiment (Fig. 2d).

Comparison between Hg1201 and YBCO [6,7] demonstrates remarkable universality (Fig. 3): (I) In both cases, the order preserves the translational symmetry of the underlying lattice, unlike conventional (1/2,1/2,0) type antiferromagnetism; (II) The magnetic scattering develops below a temperature which coincides with $T^*$ determined from DC transport, suggesting that the order involves both magnetic and charge degrees of freedom; (III) The magnetic signal is of comparable strength for the two compounds, it is strongest in very underdoped samples, and the transition appears to be continuous. We note that the effect in the most underdoped sample B' is very strong, corresponding to $\sim 0.2$ $\mu_B$ per unit cell in a naïve picture of spin-based moments, and that the present data statistics do not allow a reliable determination of the order parameter critical exponent; (IV) Using previous estimates for the doping dependence of $T_c$ [26,27], the ordering temperatures for both systems fall onto the same line. Linear extrapolation suggests that $T_{mag}$ approaches zero close to the value $p_c$=0.19, which has been argued to be the location of a quantum critical point [5]. Alternatively, rescaling $T_c(p)$ for Hg1201 to the curve for YBCO shifts $T_{mag}(p)$ to higher hole concentrations and leads to an apparent disappearance of the magnetic signal near $p$=0.15 for both systems (see supplementary information). Both the linear trend and the value of $p_c$ are consistent with new polar Kerr effect results for YBCO that also indicate the existence of a phase with broken time reversal symmetry, although with ordering temperatures that are systematically lower [28]; (V) In both cases, the moment does not lie along the c-axis, but rather has a considerable in-plane component. For sample B, the (101) intensity measured with $\mathbf{P} \perp \mathbf{Q}$ ($\mathbf{P}$ in the scattering plane) is approximately 65% of that for $\mathbf{P} // \mathbf{Q}$ (Fig. 2c). Noting that polarized neutron diffraction in the SF channel probes the component of the magnetic moments perpendicular to both $\mathbf{P}$ and $\mathbf{Q}$, and that in the former geometry $\mathbf{P}$ makes a relatively small angle with the c-axis, this suggests a non-negligible ab-plane component of the measured moment.

The observation of magnetic Bragg scattering at $\mathbf{Q}$=(101) is consistent with an even number of moments per unit cell with zero net moment. Magnetic order involving spin moments on the planar oxygen atoms (Fig. 1f) could, in principle, preserve the translational invariance of the underlying crystal lattice [6]. However, such order would be difficult to reconcile with the unusual moment direction and, as discussed below, with the observed strong $Q$-dependence. Moreover, it should be discernable with NMR, yet no such evidence has been reported [24]. Instead, it appears likely that the novel state arises from circulating charge currents [11,12]. The experiments for Hg1201 and YBCO [6,7] are qualitatively consistent with magnetism due to two counter-circulating charge current loops per CuO$_2$ plaquette (Fig. 1g), but since the theory involves the planar oxygen $p$- and copper $d$-orbitals, it predicts a magnetic moment along the $c$-axis, which cannot explain the in-plane component found experimentally. In order to explain the unusual moment direction and the tiny ferromagnetic component observed by the polar Kerr effect [28], it has been proposed that the relatively low structural symmetry of YBCO will lead to spin-orbit coupling that causes spin order to accompany planar loop-current order [29]. However, such spin-orbit coupling is expected to be absent in Hg1201, which possesses a high tetragonal structural symmetry in which the planar Cu and O sites are centers of inversion. The presence of significant O spin moments is furthermore inconsistent with the narrow [17]O NMR linewidth [24].

The correct description might be a variant of the proposed phase, with orbital-current loops that involve the apical oxygens, but without current flow through the copper site (Fig. 1h). The tilt angle spanned by the CuO$_2$ plane and the 'oxygen triangles' is about 64° for Hg1201 and 59° for YBCO, consistent with the fact that a large portion of the total signal ($\mathbf{P}//\mathbf{Q}$) is distributed in the $\mathbf{P} \perp \mathbf{Q}$ ($\mathbf{P}$ in the scattering plane) geometry. These observations are further supported by the data in Fig. 2g, which demonstrate that the magnetic signal is even stronger at $\mathbf{Q}$=(100) than at (101), while at the (201) and (102) reflections no signal was discerned. With good agreement with the results for YBCO [6], this trend is consistent with the general expectation that magnetic signal decreases with increasing $Q$. However, the dramatic decrease of the intensity with increasing $Q$ implies that, distinct from conventional spin order, the magnetic density has a large spatial extent, consistent with a picture of extended spontaneous orbital currents within the unit cell. Moreover, the currents cannot be confined to the CuO$_2$ planes (Fig. 1g), since this would not lead to a strong $L$-dependence. Very recent theoretical work on extended two-dimensional Hubbard models including apical oxygen orbitals support this picture [14].

The maximum $T_c$ occurs close to where the experiment fails to discern a magnetic signal, and it appears likely that the order competes with the superconductivity. One intriguing possibility is that the fluctuations associated with an underlying quantum critical point are directly responsible for the appearance of superconductivity and the unusual normal state properties, such as the linear resistivity found up to remarkably high temperatures.

**Acknowledgements**

We thank Henri Alloul and Chandra Varma for helpful comments. The work at Stanford University was supported by grants from the US Department of Energy and the National Science Foundation.

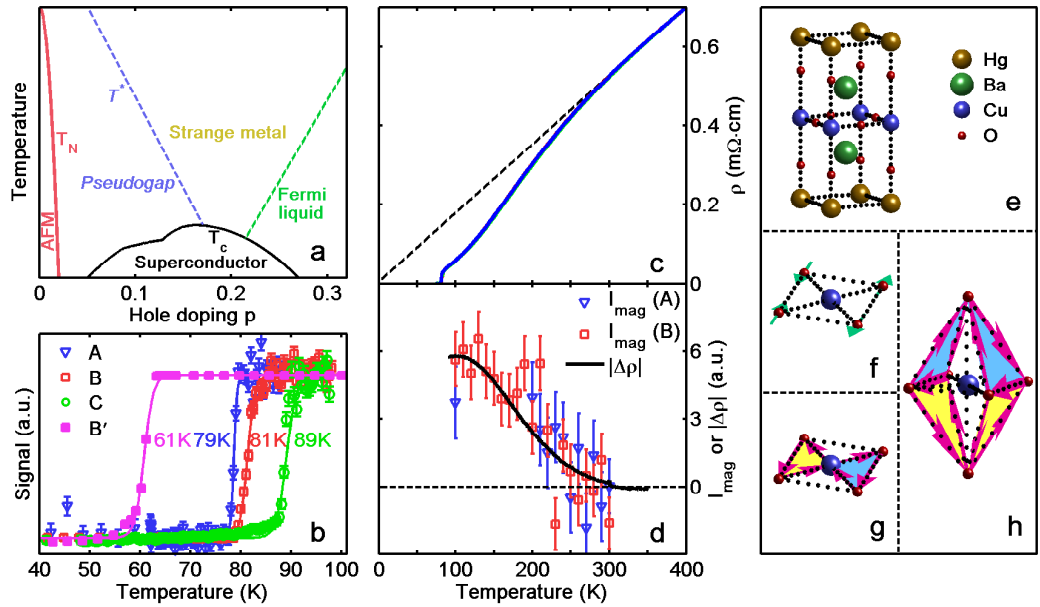

**Figure 1 | Pseudogap in underdoped Hg1201. a,** Schematic phase diagram of hole-doped high-$T_c$ superconductors; **b,** determination of $T_c$ by SQUID magnetometry (B') and neutron depolarization (A,B,C) of the four underdoped Hg1201 samples studied in this work. Quoted $T_c$ values are transition mid-points. Sample B' was obtained from sample B (mass ~ 600mg, as-grown) upon annealing in partial vacuum (0.1 Torr) at 450 ºC. Samples A (150mg) and C (1.2g) are as-grown. Typical sample mosaic is less than 0.5º (FWHM). The neutron-depolarization effect was used to measure the bulk $T_c$ whenever possible. A guide field of ~10 Oe was applied along the beam path. After the samples were cooled below $T_c$ and contained trapped vortices, the guide field at the sample position was turned by 90º, resulting in an abrupt magnetic field change at the sample surface, which can be observed as a decrease in *FR*. In several cases, $T_c$ was verified using conventional magnetic susceptometry; **c,** temperature dependence of resistivity for a separate crystal ($T_c$=79K). The resistivity measurement employed the standard four-probe method, with electrical contacts sputtered on the *ac/bc* faces of a small single crystal (contact resistance less than one ohm); **d,** deviation from linear resistivity compared with magnetic signal for samples A and B, demonstrating that the novel magnetic order is linked to the pseudogap; **e,** crystal structure of Hg1201 (dopant oxygen atoms in Hg-O layer not shown); **f-h,** simplified schematic illustrations of three ordered states that break time-reversal symmetry, but preserve translational symmetry: **f,** spin-order involving oxygen atoms; **g,** planar orbital currents; **h,** orbital currents involving apical oxygens.

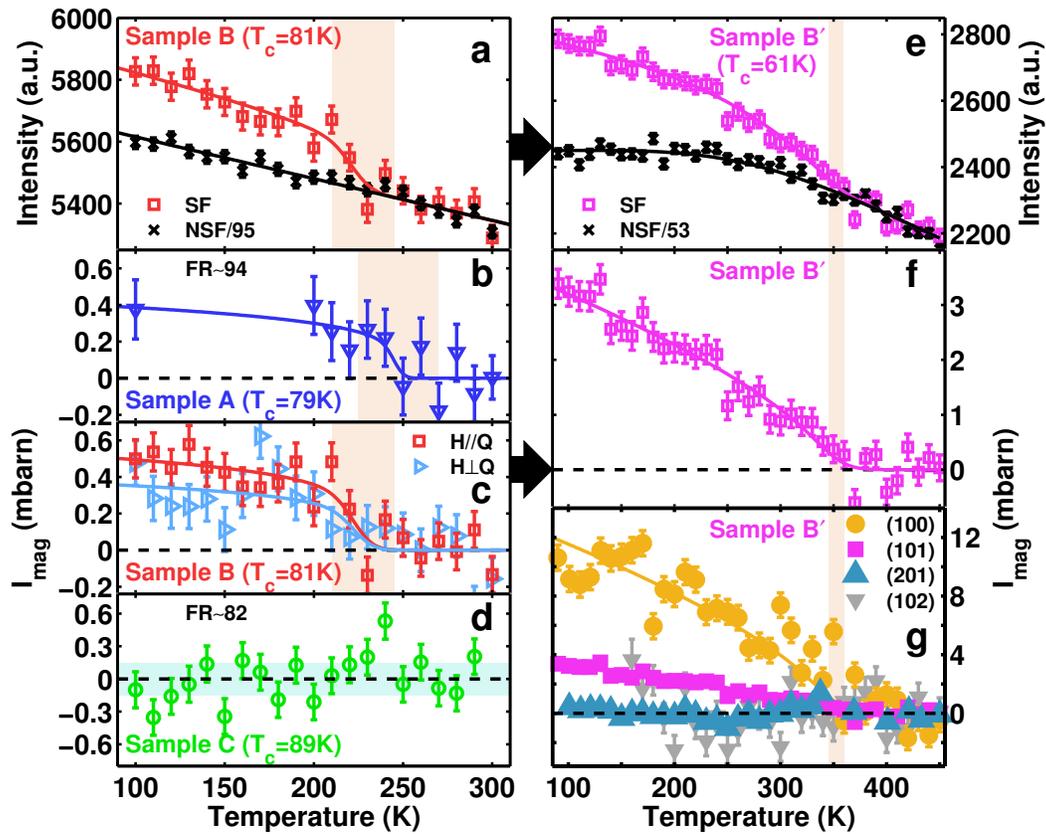

**Figure 2 | Novel magnetic order revealed by polarized-neutron diffraction.**
**a**, **e**, raw data for samples B and B'. Magnetic signal appears as additional intensity in the spin-flip (SF) channel compared to "background" intensity due to nuclear Bragg scattering. The latter is measured in the non-spin-flip (NSF) channel, but a fraction 1/*FR* is also observed in the SF geometry; **b-d**, **f**, temperature dependence of net intensity, which is obtained after the removal of the background. Conversion to absolute units is completed using the intensity of the same nuclear Bragg peak; **g**, intensity measured on different Bragg peaks in the most underdoped sample. Data in **a-f** are collected on the Bragg peak $\mathbf{Q} = (101)$, with the neutron spin parallel to $\mathbf{Q}$, and in **c** also with the neutron spin perpendicular to $\mathbf{Q}$ in the scattering plane. Vertical bands indicate the estimated onset temperatures of the magnetic signal; horizontal band in **d** represents an upper bound estimate for the intensity of sample C. Solid colored lines in **a-c**, **e-g** are guides to the eye. Horizontal arrows represent the oxygen-anneal step carried out to obtain sample B'. Error bars represent counting statistics (one standard deviation). The polarized-neutron diffraction experiments were carried out in continuous runs with minimal instrument movement. The super-mirror polarizer and Heusler analyzer were arranged such that the electromagnetic spin flipper was off when measuring the SF channel, providing maximum stability. The experiments are limited to temperatures above $T_c$, because the required neutron guide field cannot be reliably sustained in the superconducting state.

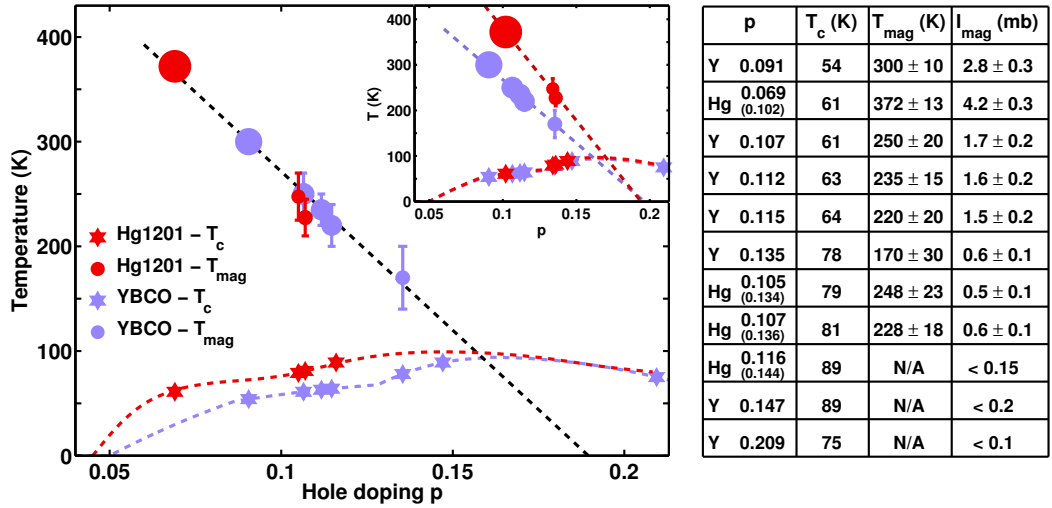

| | p | $T_c$ (K) | $T_{mag}$ (K) | $I_{mag}$ (mb) |
|---|---|---|---|---|
| Y | 0.091 | 54 | 300 ± 10 | 2.8 ± 0.3 |
| Hg | 0.069 (0.102) | 61 | 372 ± 13 | 4.2 ± 0.3 |
| Y | 0.107 | 61 | 250 ± 20 | 1.7 ± 0.2 |
| Y | 0.112 | 63 | 235 ± 15 | 1.6 ± 0.2 |
| Y | 0.115 | 64 | 220 ± 20 | 1.5 ± 0.2 |
| Y | 0.135 | 78 | 170 ± 30 | 0.6 ± 0.1 |
| Hg | 0.105 (0.134) | 79 | 248 ± 23 | 0.5 ± 0.1 |
| Hg | 0.107 (0.136) | 81 | 228 ± 18 | 0.6 ± 0.1 |
| Hg | 0.116 (0.144) | 89 | N/A | < 0.15 |
| Y | 0.147 | 89 | N/A | < 0.2 |
| Y | 0.209 | 75 | N/A | < 0.1 |

**Figure 3 | Universal pseudogap phase diagram.** In the main panel, hole doping is estimated from the $T_c(p)$ relationships (dotted lines) reported in [26,27]. $T_{mag}$ for Hg1201 and YBCO is determined in this work and in [6,7], respectively. A linear fit of $T_{mag}(p)$ to the combined data extrapolates to $T_{mag}$= 0K at $p_c$ = 0.190±0.011. Note that the value of $T_c$ is a function not only of $p$, but also of disorder [23,26], possibly leading to systematic differences in carrier concentration estimates. Furthermore, the $T_c(p)$ relationships for Hg1201 and YBCO differ below optimal doping, and the systematic deviation from a parabolic form might result from a tendency toward stripe-order formation near $p$=1/8 [26]. In the inset, $p$ for Hg1201 is estimated using the $T_c(p)$ relationship for YBCO [26], and linear extrapolation gives $p_c$ = 0.194±0.025 (Hg1201) and $p_c$ = 0.196±0.011 (YBCO). Symbols are plotted with area proportional to signal intensity estimated at $T$=0K. Samples at relatively high doping did not exhibit an observable magnetic signal (see also supplementary material for YBCO). Error bars represent the uncertainty in the estimation of $T_{mag}$. The results are summarized in the table; $p$ values for Hg1201 pertaining to the inset are given in parenthesis. For the most underdoped Hg1201 sample B', an order parameter fit was attempted, allowing for a small distribution of transition temperatures $T_{mag}$. This distribution was estimated from the superconducting transition width and the approximate slope of $T_{mag}$ vs. $T_c$. Fits to the (100) and (101) Bragg peaks were carried out to extract $T_{mag}$ and $\beta$; $\beta$ was found to be strongly dependent on the range of the fit, which was unstable for $T$>300K, and yielded $\beta$=0.18±0.13 for $T$>250K and $\beta$=0.42±0.12 for $T$>200K.

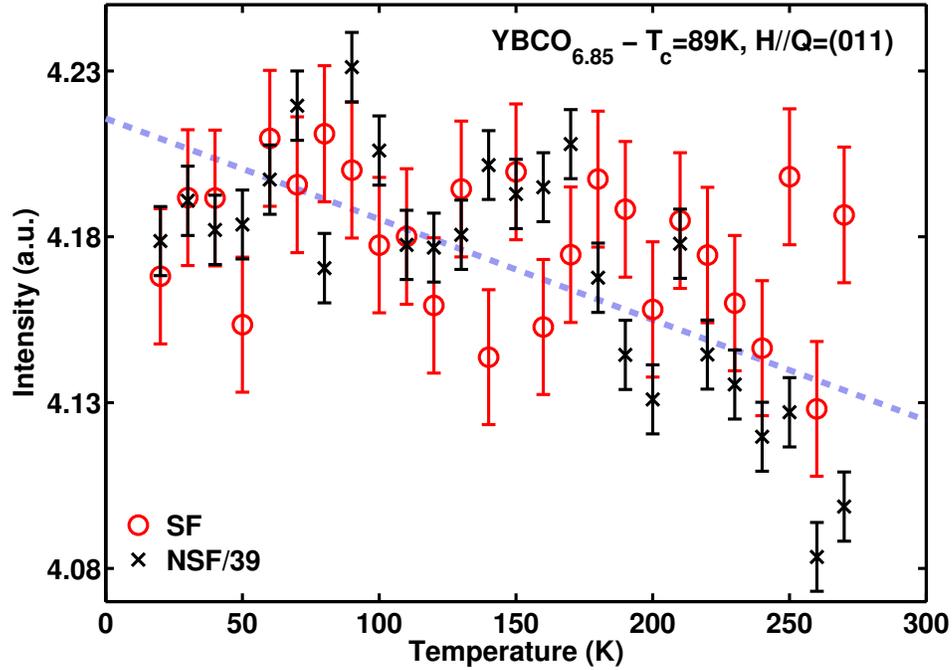

**Supplementary Information**

**Figure 1 | Absence of magnetic signal in nearly optimally-doped YBa$_2$Cu$_3$O$_{6.85}$.** A nearly optimally doped YBa$_2$Cu$_3$O$_{6.85}$ sample ($T_c$ = 89K, same as Hg1201 sample C in this work) [Bourges, P. *et al.*, Science **288**, 1234 (2000)] was studied (Fig. 1, unpublished). A comparison between the SF and NSF channels demonstrates the absence of a magnetic signal within the experimental sensitivity, with an upper bound of 0.2 mbarn on the magnetic intensity.